\documentclass[DIV=12]{article}
\usepackage[margin=1.1in]{geometry}
\usepackage{subfig}

\usepackage[dvips]{color}
\usepackage{graphicx}
\usepackage{amsmath}

\begin{document}

\title{
\noindent\hrulefill
\begin{flushleft}
{\Large \bf{Cell-type-specific neuroanatomy of  brain-wide expression of autism-related genes}}\\
\vspace{5mm}
\hrule
\vspace{6mm}
{\normalsize{Pascal Grange\\
Department of mathematical sciences\\
Xi'An Jiaotong-Liverpool University, Suzhou 215123, Jiangsu, China\\
{\ttfamily{pascal.grange@xjtlu.edu.cn}}\\
\vspace{6mm}
Idan Menashe\\
Department of Public Health, Faculty of Health Sciences\\
Ben-Gurion University of the Negev, P.O.Box 653, Beer-Sheva 84105, Israel\\
\vspace{6mm}
Michael Hawrylycz\\
\vspace{-3mm}
Allen Institute for Brain Science, Seattle, WA 98103, United States}}
\end{flushleft}}
\date{}
\author{}
\maketitle

\begin{abstract}
 Two cliques  of genes identified computationally for their high co-expression in the  mouse brain according 
 to the Allen Brain Atlas, and for their enrichment in genes related to autism spectrum disorder, have recently 
 been shown to be highly co-expressed in the cerebellar cortex, compared to what could be expected 
 by chance. Moreover, the expression of these cliques of genes is not homogeneous across the cerebellum,
 and it has been noted that their gene expression pattern seems to highlight the granular layer. However,
 this observation was only made by eye, and recent advances in computational neuroanatomy
 allow to rank cell types in the mouse brain (characterized by their transcriptome profiles) according to the similarity 
 between their density profiles and the expression profiles of the cliques. We establish by Monte Carlo simulation that 
 with probability at least 99\%, the expression profiles of the two cliques are more similar to the density profile of granule cells
  than 99\% of the expression of cliques containing the same number of genes (Purkinje cells also score above 99\% in one of the cliques). 
 Thresholding the expression profiles shows that the signal is more intense in the granular layer.

\end{abstract}

\vspace{8mm}


{\bf{Acronyms.}} ASD: autism spectrum disorder; ABA: Allen Brain Atlas; ISH: {\emph{in situ}} hybridization; ARA: Allen Reference Atlas; 
CDF: cumulative distribution function.\\
\vspace{6mm}

\tableofcontents

\section{Introduction}

 The neuroanatomical structures underlying autism spectrum disorder (ASD) traits are 
the subject of intense research efforts, as ASD is one of the most prevalent
and highly heritable neurodevelopmental disorders in
 humans \cite{autismLevy,epidemiologyLord,epidemiologyNewSchaffer, neuroanatomyAmaral}.
 A good source of data for such studies is the Allen Brain Atlas (ABA) of the 
 adult mouse \cite{AllenAtlasMol, AllenFiveYears,corrStructureAllen, AllenGenome,images,BrainAtlasInsights,digitalAtlasing,NeuroBlast,neufoAllen,ISHVsMicroarray}, which consists of 
  thousands of brain-wide {\emph{in situ}} 
hybridization (ISH) gene-expression profiles, 
 co-registered to the Allen Reference Atlas (ARA) \cite{ARA}.
  Recently, we used the ABA to examine the spatial
co-expression characteristics of genes associated with autism
susceptibility \cite{autismCoExpr}. We identified two networks of  co-expressed
genes that are enriched with autism genes and
 which are significantly overexpressed
in the cerebellar cortex.\\


These results added to the mounting evidence of the involvement of the cerebellum in
 autism \cite{refCerebellum,refPurkinje}. However, the rich internal structure of cerebellum requires a 
further investigation of the specific cerebellar regions or cell types associated with ASD.\\

In our paper \cite{autismCoExpr} we indicated that the two cliques of co-expressed autism genes appear to be overexpressed in the granular layer of the cerebellum. However, this observation was based on visual comparison of the expression patterns 
of the genes in these two cliques to sections of the estimated density patterns of cell types,
 which at the time were available as preprint from \cite{suppl1}. This approach by mere visual inspection is far from satisfactory since it does not make use of the full computational power of the ABA \cite{methodsPaper,qbCoExpression,markerGenes,BGEAManual}. Moreover, post-mortem studies of brains of autistic patients \cite{postMortem} have shown alterations in the Purkinje layer of the cerebellum, rather than in the granule cells.\\

In the present study we re-examine the two cliques discovered in \cite{autismCoExpr} using recent developments of computational neuroanatomy relating cell-type-specificity of gene expression to neuroanatomy 
  based on the ABA. For example, in \cite{cellTypeBased,suppl1,suppl2}, the region-specificity of 64 cell types (collated from \cite{foreBrainTaxonomy,ChungCells,ArlottaCells,RossnerCells,HeimanCells,CahoyCells,DoyleCells,OkatyCells,OkatyComparison})
 was estimated using a linear mathematical model, which amounts to decomposing the 
 gene expression in the ABA over a set of measured cell-type-specific transcriptomes (see also \cite{TanFrenchPavlidis,KoCellTypes} for 
 cell-type-specific analyses of the ABA, and \cite{microarrayLupus} for 
 a similar mathematical approach in the context  of blood cells). We extend the 
 Monte Carlo methods (developed in \cite{autismCoExpr} to estimate the probability
 of co-expression among a set of genes) to the comparison between 
 the expression of a set of genes and the spatial density profile of a cell type.
 This allows to estimate not only the probability of the similarity between a gene 
 clique and the granular layer, but also the probabilities of similarity to
  the spatial distributions of all cell types considered in \cite{cellTypeBased}.




\section{Methods}

{\bf{Cliques of genes.}} We re-examine the brain-wide expression 
 profiles of the two cliques
 $\mathcal{C}_1$ and $\mathcal{C}_2$  of genes identified in \cite{autismCoExpr} based on
 their exceptional co-expression profiles, which
 consist of 33 and 6 genes respectively:\\
\begin{equation}
  \begin{array}{ll}  
\mathcal{C}_1 &= \{ Astn2, Dpp6, Galnt13, Ptchd1,
        Trim3, Slc12a3,\\
      & Pltp, Mpp3, Darc, B230317C12Rik, Pla2g7, Syt2,\\
      & Edg1, Cnr1, 0610007P14Rik, Socs5,
        Atp1a1, Chgb,\\
     & Car4, Pcbp4, Syne1, Camk2d, Slc6a1,C230009H10Rik,\\
     &LOC434631, Prpf38b, D530033C11Rik, Coro2b, Tmem109, Daam2,\\
      &  Gpr37l1, BC060632, Grm4\},
\end{array}
\end{equation}
\begin{equation}
 \begin{array}{ll}  
\mathcal{C}_2 = \{Rims3, Astn2,
        B230308C24Rik\ast, LOC434631, 4933417O08Rik, Car10\}.
\end{array}
\end{equation}\\
They both contain genes from the AutRefDB database (\cite{AutRefDB1,AutRefDB2}) of ASD-related genes({\emph{Ptchd1, Galnt13, Dpp6}} and {\emph{Astn2}} for the first clique,
 {\emph{Astn2}} and {\emph{Rims3}} for the second).\\

{\bf{Gene expression energies from the Allen Brain Atlas.}} The adult mouse brain is partitioned into $V=49,742$ cubic voxels of side 200 microns, to which ISH data are registered \cite{ARA,AllenGenome} for thousands of genes.
For computational purposes, these gene-expression data can be arranged into 
 a voxel-by-gene matrix. For a cubic voxel labeled $v$, the {\it{expression energy}} of the gene $g$ is a
weighted sum of the greyscale-value intensities evaluated at the
pixels intersecting the voxel:
\begin{equation}
E(v,g) = {\mathrm{expression\;energy\;of\;gene\;labeled\;}}g\;{\mathrm{in\;voxel\;labeled\;}}v,
\label{ExpressionEnergy}
\end{equation}
  The present analysis is restricted to the coronal atlas, 
 as in \cite{qbCoExpression,BGEAManual,autismCoExpr}, for which the entire mouse brain was processed in 
 the ABA pipeline (whereas only the left hemisphere was processed for the sagittal atlas)\\

{\bf{Cell-type-specific microarray data and estimated cell-type-specific density profiles .}} The
 cell-type-specific microarray reads collated in \cite{OkatyComparison} from the studies \cite{foreBrainTaxonomy,ChungCells,ArlottaCells,RossnerCells,HeimanCells,CahoyCells,DoyleCells,OkatyCells} (for $T=64$
 different cell-type-specific samples) are arranged in a type-by-gene matrix denoted by $C$, such that
\begin{equation}
C(t,g) = {\mathrm{expression\;of\;gene\;labeled\;}}g\;{\mathrm{in\;cell\;type\;labeled\;}}t,
\label{typeByGene}
\end{equation} 
 and the columns are arranged in the same order as in the matrix $E$ of expression energies defined in Eq. \ref{ExpressionEnergy}.
  In \cite{cellTypeBased}, we proposed a simple linear model for a voxel-based gene-expression atlas in terms
of the transcriptome profiles of individual cell types and their spatial densities:
\begin{equation}
E(v,g) = \sum_t  \rho_t(v)C( t,g)  + {\mathrm{Residual}}(v,g),
\label{modelEquation}
\end{equation}
where the index $t$ denotes the $t$-th cell type, with density profile $\rho_t(v)$ at voxel labeled $v$. The values
 of the cell-type-specific density profiles were computed in \cite{cellTypeBased} by minimizing the value
 of the residual term over all the (positive) density profiles, which amounts to solving a quadratic
 optimization problem at each voxel.\\

{\bf{Measure of similarity between gene-expression patterns and cell-type-specific density patterns.}} 
 The quantitative study of spatial co-expression of genes conducted in \cite{autismCoExpr} combines the columns
 of the matrix of gene-expression energies (Eq. \ref{ExpressionEnergy}) by computing the cosine similarities
  of all pairs of genes in the cliques  $\mathcal{C}_1$ 
 and $\mathcal{C}_2$. These cosine similarities are then compared to those obtained from random sets
 of genes containing the same numbers of elements as $\mathcal{C}_1$ 
 and $\mathcal{C}_2$ respectively. This technique can be adapted to compare brain-wide 
gene-expression profiles to the spatial density of cell types,
 simply by considering cosine similarities between gene-expression profiles and cell-type-specific density profiles.\\
Given a set $\mathcal{G}$ of genes from the coronal ABA (selected either computationally based on their co-expression
 properties, or based on curation of the biomedical literature, which in the present case means $\mathcal{G} = \mathcal{C}_1$ or $\mathcal{G}=\mathcal{C}_2$), we can 
 compute the sum of their expression profiles:
 \begin{equation}
  E^{\mathcal{G}}( v ) = \sum_{i = 1}^{| \mathcal{G}|} E(v,g_i),
 \end{equation}
 where $g_i$ is the column index in the matrix of expression energies (Eq. \ref{ExpressionEnergy})
 corresponding to the $i$-th gene in the set $\mathcal{G}$. 
 The quantity  $E^{\mathcal{G}}$ is an element of $\mathbf{R}_+^V$, just as the estimated brain-wide density profile
 of a cell type. We can therefore estimate  the similarity between  $E^{\mathcal{G}}$ and the 
 density of cell type labeled $t$ by computing the cosine similarity
\begin{equation}
\psi(\mathcal{G},t) = \frac{\sum_{v=1}^VE^{\mathcal{G}}( v ) \rho_t(v)}{\sqrt{\sum_{u=1}^V E^{\mathcal{G}}( u )^2 } \sqrt{\sum_{w=1}^V \rho_t( w )^2 }},
 \label{cosineSimilarity}
\end{equation}
 which is between 0 and 1 by construction.\\


{\bf{Statistical significance of the similarity between expression patterns and density patterns of cell types.}} 
   Furthermore, for a fixed cell type, we can estimate how exceptional 
the similarity $\psi(\mathcal{G},t)$ is, compared to what would be expected from random sets 
 of $|\mathcal{G}|$  genes  drawn from the coronal ABA. This is a finite problem,
 but it becomes hugely complex in a regime where $|\mathcal{G}|$ is relatively
 large but still small compared to the size of the entire atlas (which is the case for both cliques in the present study). We can  
 take a Monte Carlo approach, draw $R$ random sets of $|\mathcal{G}|$ genes
 and simulate the cumulative distribution function (CDF) of the cosine similarity\footnote{or any other measure
 of similarity}
  between a random set of $|\mathcal{G}|$ genes and
 the density profile of cell-type labeled $t$ (this function depends only on the cell type and on the
 number of genes $|\mathcal{G}|$, we can denote it by ${\sc{\mathrm{CDF}}}_{t,|\mathcal{G}|}$). If we denote by $\mathcal{G}_1$, $\mathcal{G}_2$,...
 $\mathcal{G}_R$  random subsets of $[1..G]$ (drawn without repetition), we obtain an estimate an 
 estimate of the CDF that reads as:
\begin{equation}
{\sc{\mathrm{CDF}}}_{t,|\mathcal{G}|,R}(x) := \frac{1}{R}\sum_{r=1}^R\mathbf{1}\left(   \psi(\mathcal{G}_r,t)   \geq  x   \right)\;\longrightarrow_{R\infty} {\sc{\mathrm{CDF}}_{t,|\mathcal{G}|}(x)}.
 \label{estimateCDF}
\end{equation}
Moreover, the probability of obtaining
 a similarity to $\rho_t$ larger than a threshold $\psi(\mathcal{G},t)$ is estimated by:
\begin{equation}
\mathcal{P}_R(\mathcal{G}, t) := {\sc{\mathrm{CDF}}}_{t,|\mathcal{G}|,R}(\psi(\mathcal{G},t)  ) =\frac{1}{R}\sum_{r=1}^R\mathbf{1}\left(   \psi(\mathcal{G}_r,t)   \geq  \psi(\mathcal{G},t)   \right).
 \label{estimateP}
\end{equation}

The precision of our estimates depends on the value of $R$. We can use Hoeffding's inequality to compute a lower bound  
 on the number $R$ of  random draws required to estimate the probability 
 of being within a known error from the true CDF. As we are estimating the probability 
 of having larger cosine similarity than expected by chance by summing 
 $R$ Bernoulli variables (Eq. \ref{estimateP}),
  Hoeffding's inequality (see \cite{ML} for instance) states that for any $\tau$, the probability of missing the 
 true value of the probability  $\mathcal{P}(\mathcal{G}, t)$ by $\tau$ is bounded in terms of $\tau$ and the number of random 
 draws $R$ as follows:
\begin{equation}
P(|\mathcal{P}_R(\mathcal{G}, t) - \mathcal{P}(\mathcal{G}, t)|\geq \tau ) \leq \exp( -2R\tau^2).
\label{Hoeffding}
\end{equation}
 For instance, taking $\tau = 0.01$ and $R = 26,500$ leads to a value of $0.01$ for the
 bound on the r.h.s. of the inequality \ref{Hoeffding}, so it is enough to draw this number
 of random sets of genes to obtain an estimator within 1 percent of the true probabilities,
 with probability at least 99 percent.\\

  Having conducted the simulation of the distribution of cosine similarities 
 for a choice of $R$  based on
 Hoeffding's inequality, we can rank cell types for a fixed clique $\mathcal{G}$
 by decreasing values of statistical significance:\\
\begin{equation}
 {\mathcal{P}}_R(  \mathcal{G}, t_\mathcal{G}( 1 ) )\geq {\mathcal{P}}(  \mathcal{G}, t_\mathcal{G}( 2 ) )
  \geq ... \geq    {\mathcal{P}}_R(\mathcal{G}, t_\mathcal{G}( T ) ).
 \label{rankTypes}
\end{equation}

{\bf{Similarity between thresholded gene-expression energies and cell-type-specific densities.}} 
 Given that  the expression profiles of the cliques is much less sparse than any of the densities
 of cell types estimated in \cite{cellTypeBased}, the genes in the cliques must be expressed 
 in several different cell types, but there are large differences in expression between cortical voxels
 and cerebellar voxels for instance, and also within the cerebellar cortex (see Fig. \ref{figHeatMaps}a,b). 
 We propose to threshold brain-wide expression 
 profile of each clique, and to recompute the cosine similarities with 
 density profiles, in order to discover which 
 neuroanatomical cell-type-specific patterns are highlighted with more intensity. 
 If the profile of a given cell type is highlighted by a given clique, when the threshold
 grows from zero to low values of the threshold,
 the cosine similarity is expected to grow, since many voxels with 
 low values of expression energy, that penalize
 the cosine similarity to  the cell type, are put to zero by the threshold. Let us denote by
 $\tau$ the value of the threshold. We can compute the thresholded
  expression energies of the cliques and cosine similarities as follows:
\begin{equation}
 E^{\mathcal{G}}_\tau(v) = E^{\mathcal{G}}(v)\mathbf{1}\left( E^{\mathcal{G}}(v) \geq \tau  \right),
\label{EThresh}
\end{equation}
\begin{equation}
\psi_\tau(\mathcal{G},t) = \frac{\sum_{v=1}^VE^{\mathcal{G}}_\tau( v ) \rho_t(v)}{\sqrt{\sum_{u=1}^V E_\tau^{\mathcal{G}}( u )^2 } \sqrt{\sum_{w=1}^V \rho_t( w )^2 }},
\label{coExprThresh}
\end{equation}
At very large values of the threshold, expression energies 
 are going to be put to zero everywhere, 
 and the cosine similarities decrease to zero.
 So the cosine similarity between the expression of the two gene cliques 
  and the cell types they highlight are expected to exhibit peaks when plotted as a function
 of the threshold. The higher the peak, and the higher the corresponding value of the threshold, the 
 more intensely the cell type is highlighted.

\section{Results}



 We computed the cosine similarities between the expression profiles of the two cliques
 ${\mathcal{C}}_1$ and ${\mathcal{C}}_2$ and the density profiles
 of the $T=64$ cell types estimated in \cite{cellTypeBased}, 
 using Eq. \ref{cosineSimilarity}.
  For each cell type, we computed the   probabilities $\mathcal{P}_R({\mathcal{C}}_1,t)$
 and $\mathcal{P}_R({\mathcal{C}}_2,t)$ for $R=27,000$. 
Tables \ref{tableResultsCliques} show 
 the cell types for which the cosine similarity is larger than 10\%, ordered by decreasing values 
 of statistical significance.
 For both cliques, granule cells (labeled $t=20$) and Purkinje cells (labeled $t=1$),
 have the highest value of $P_R$ (more than 99\% for both cliques in the case of granule cells).
 For each of the two cliques, one more cell type has a value of $P_R$ larger than 
 80\% (mature oligodendrocytes, labeled $t=21$, in the case of ${\mathcal{C}}_1$,
 pyramidal neurons, labeled $t = 46$, in the case of ${\mathcal{C}}_2$).
 The statistical significance (i.e. the value of ${\mathcal{P}}_R$) drops sharply after the third rank for both cliques.
  Our computational analysis therefore returns a list of four cell types to which at least one of
  the two cliques in this study are more similar than at least 80\% of the sets of 
 genes of the same cardinality as the cliques.\\

  Figure 1 shows heat maps of the expression profiles of the 
  two cliques and of the density profiles of these four cell types. 
    The expression profiles of both cliques highlight the cerebellum,
 but they are non-zero in many more voxels than any 
 of the densities of cell types illustrated in Fig. 1c1--c4.
   These densities are highly concentrated in the cerebellum
 (indeed the corresponding cell-type-specific samples were extracted from the cerebellum, see \cite{RossnerCells} for Purkinje cells, see\cite{DoyleCells} for granule cells labeled and mature oligodendrocytes),
 with the exception of the pyramidal neurons (labeled $t=46$) which are highly localized in the cerebral cortex (the corresponding  cell-type-specific samples were extracted from the layer 5 of the cerebral cortex, see \cite{foreBrainTaxonomy}).\\

\begin{table}
\caption{{\bf{Tables of cell types  ordered by decreasing values of statistical significance (see Eq. \ref{rankTypes}), measured by the probability $\mathcal{P}_R(\mathcal{G}, t)$,
  for $R=27,000$. Only cell types for which the cosine similarity is larger than 10\% are shown.}} (a) For clique 1, $\mathcal{G}=\mathcal{C}_1$. 
(b)  For clique 2 $\mathcal{G}=\mathcal{C}_2$. The top three cell types for each of the cliques have a value of  $\mathcal{P}_R(\mathcal{G}, t)$
 higher than $80\%$, and together  they consist of four distinct cell types, whose density profiles are illustrated in Fig. 1c(1--4).}
\centering
    \subfloat[\label{subfig-1:dummy}]{
      \begin{tabular}{|l|l|l|l|l|}
\hline
\textbf{Cell type}&\textbf{{\bf{Rank by significance, $t_{\mathcal{C}_1}^{-1}(t)$}}}&\textbf{Index $t$}&\textbf{$\mathcal{P}_R(\mathcal{C}_1, t)$, (\%)}&\textbf{$\psi(\mathcal{C}_1,t)$, (\%)}\\\hline
\tiny{Purkinje Cells}&1&1&100&45.9\\\hline
\tiny{Granule Cells}&2&20&100&42.4\\\hline
\tiny{Mature Oligodendrocytes}&3&21&99.5&12.7\\\hline
\tiny{GABAergic Interneurons, PV+}&4&64&38.4&35.3\\\hline
\tiny{GABAergic Interneurons, PV+}&5&59&37.6&11.3\\\hline
\tiny{GABAergic Interneurons, SST+}&6&57&36.1&22.1\\\hline
\tiny{GABAergic Interneurons, SST+}&7&56&34.8&16\\\hline
\end{tabular}

    }
    \hfill
    \subfloat[\label{subfig-2:dummy}]{
      \begin{tabular}{|l|l|l|l|l|}
\hline
\textbf{Cell type}&\textbf{{\bf{Rank by significance, $t_{\mathcal{C}_2}^{-1}(t)$}}}&\textbf{Index $t$}&\textbf{$\mathcal{P}_R(\mathcal{C}_2, t)$, (\%)}&\textbf{$\psi(\mathcal{C}_2,t)$, (\%)}\\\hline
\tiny{Granule Cells}&1&20&99.4&46.1\\\hline
\tiny{Purkinje Cells}&2&1&97.8&42.5\\\hline
\tiny{Pyramidal Neurons}&3&46&81.7&47.1\\\hline
\tiny{Mature Oligodendrocytes}&4&21&72.6&10.2\\\hline
\tiny{GABAergic Interneurons, PV+}&5&59&67.2&12.2\\\hline
\end{tabular}

    }
 \label{tableResultsCliques}
\end{table}

 \begin{figure}[!ht]
\centering
    \subfloat[\label{subfig-1:dummy}]{%
      \includegraphics[width=0.9\textwidth]{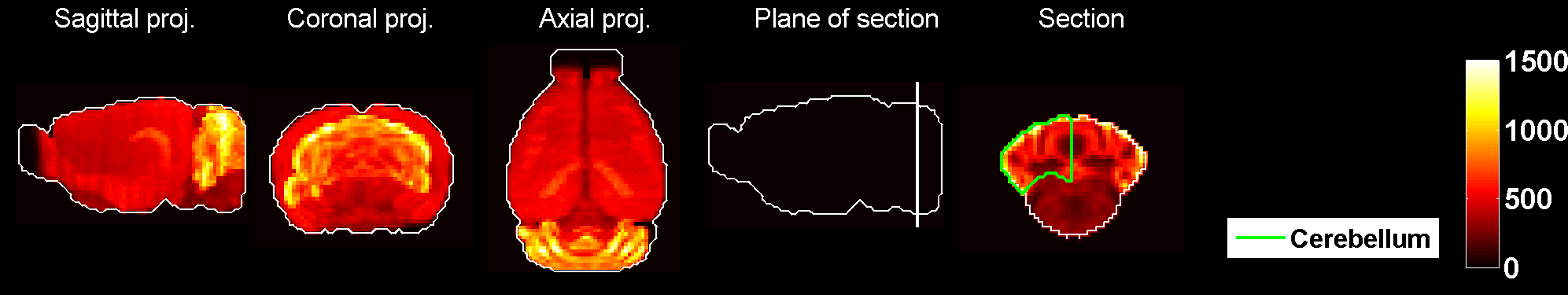}
    }
    \hfill
    \subfloat[\label{subfig-2:dummy}]{%
      \includegraphics[width=0.9\textwidth]{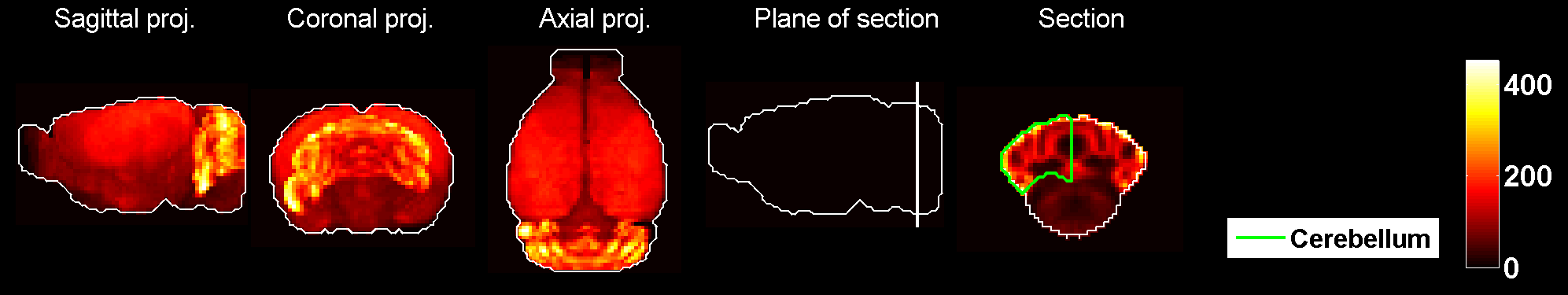}
    }
    \hfill
   \subfloat[\label{subfig-2:dummy}]{%
      \includegraphics[width=0.9\textwidth]{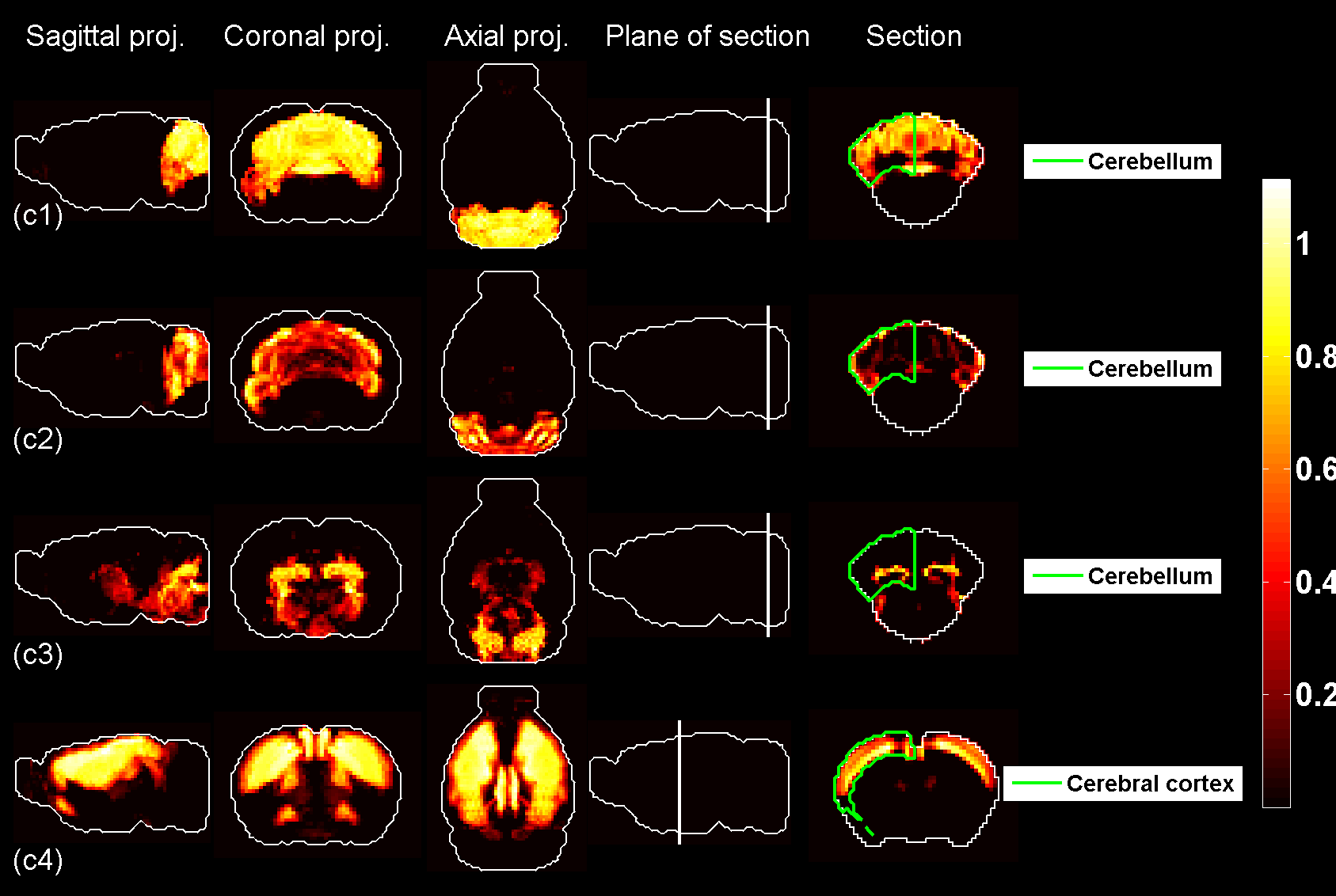}
    }
    \caption{{\bf{Heat maps of gene-expression of gene cliques, and of density profiles of cell types.}} (a) Heat map of the sum of expression energies of the 33 genes 
 in the clique ${\mathcal{C}}_1$. (b) Heat map of the sum of expression energies of the 33 genes 
 in the clique ${\mathcal{C}}_2$. (c) Heat maps of brain-wide densities (denoted by $\rho_t$ for cell type labeled $t$) of cell types estimated based on the model of Eq. \ref{modelEquation}, for Purkinje cells (c1, labeled $t=1$), granule cells (c2, labeled $t=20$), cerebellar mature oligodendrocytes (c3, labeled $t=21$), and cortical pyramidal neurons extracted from layer 5 (c4, labeled $t=46$).   These four cell types are the ones that are ranked the most highly by statistical significance of similarity to either of the cliques $\mathcal{C}_1$ and $\mathcal{C}_2$.}
    \label{figHeatMaps}
  \end{figure}



\begin{figure}[!ht]
\centering
    \subfloat[\label{subfig-1:dummy}]{%
      \includegraphics[width=0.9\textwidth]{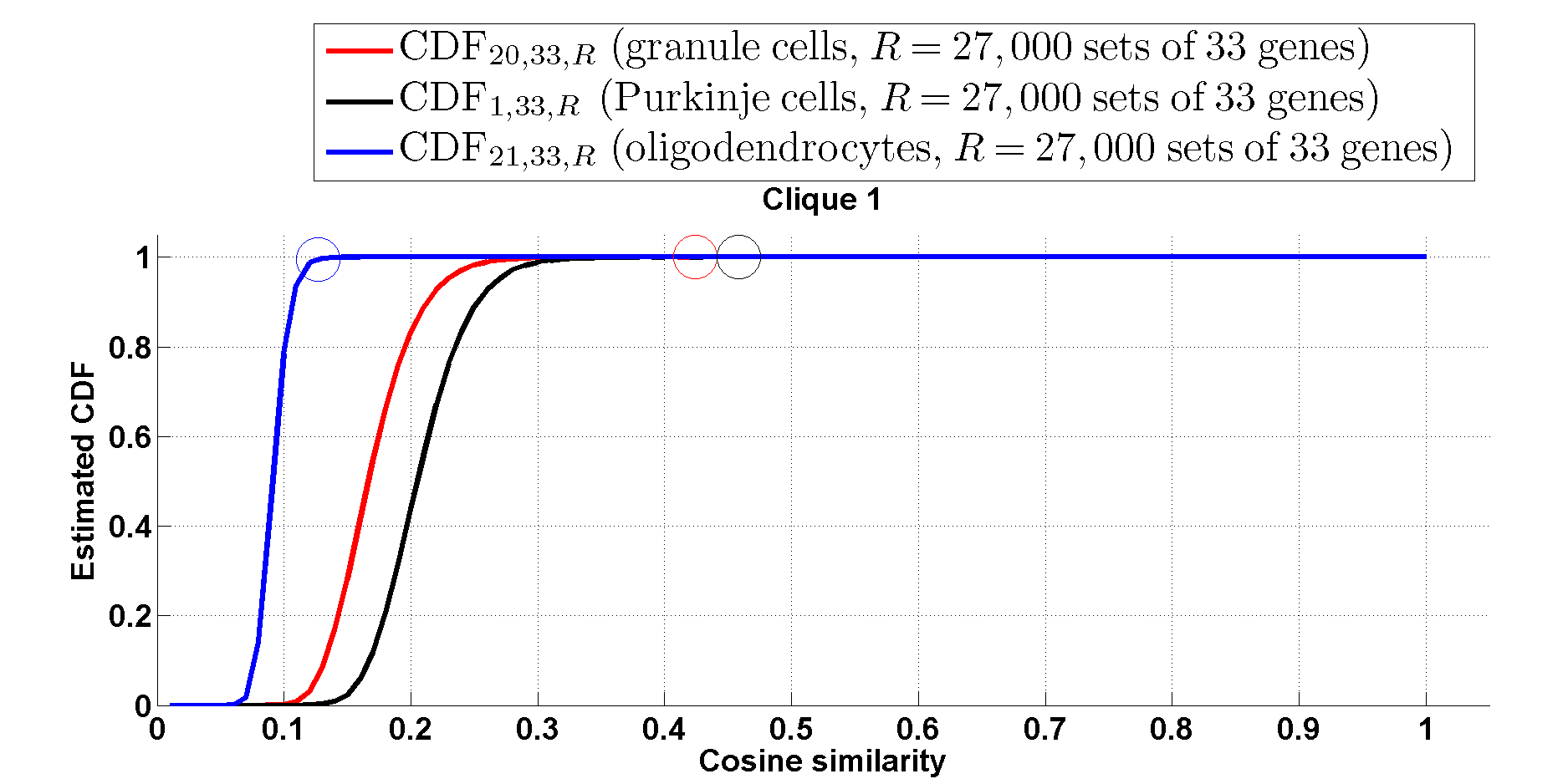}
    }
    \hfill
    \subfloat[\label{subfig-2:dummy}]{%
      \includegraphics[width=0.9\textwidth]{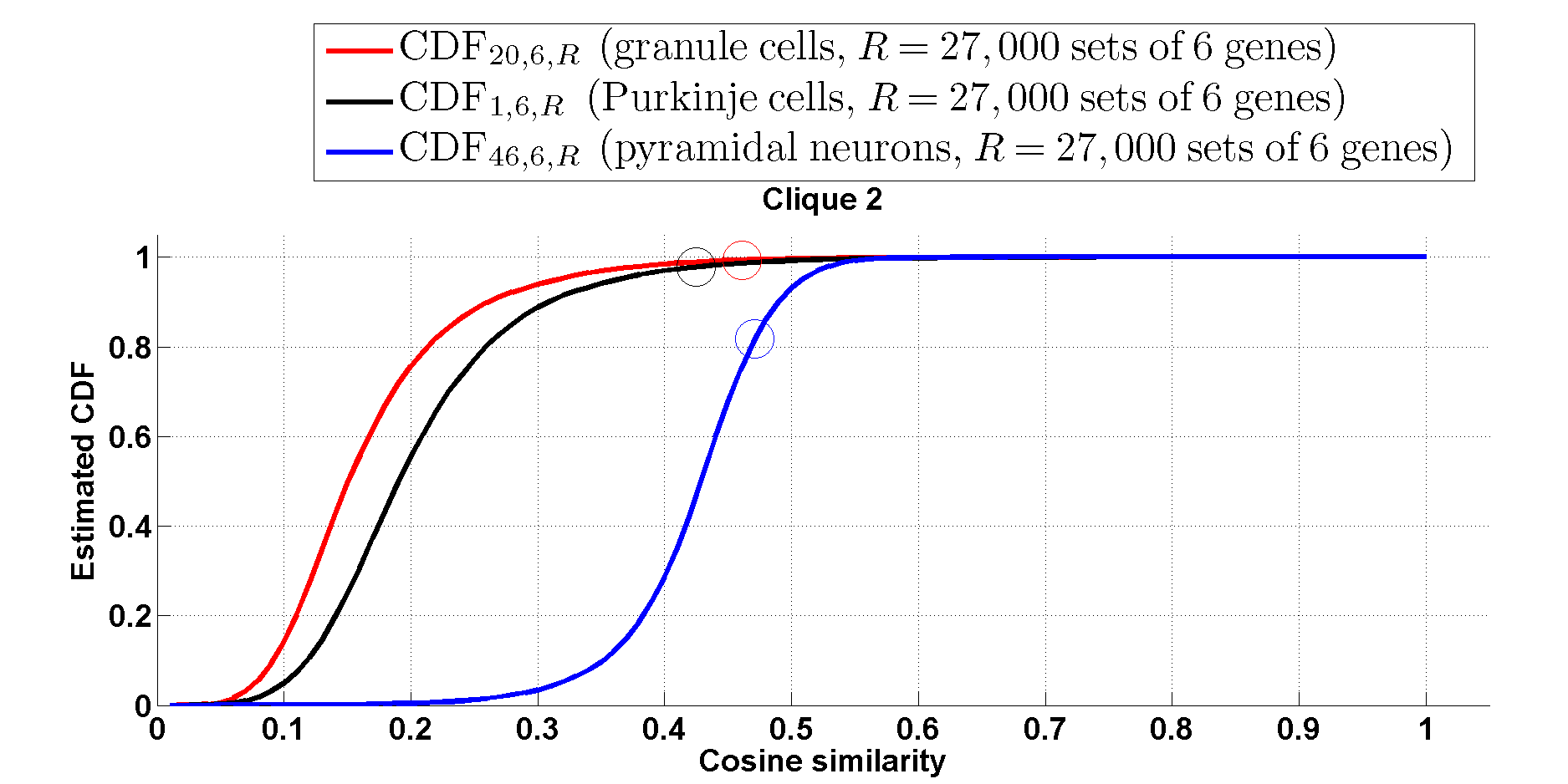}
    }
    \caption{Simulated cumulative distribution functions (CDFs) of cosine similarities between gene-expression 
 of cliques and the estimated density profile of the cell types whose density most resembles 
  one or two of the two cliques (granule cells and Purkinje cells for both cliques, along with mature oligodendrocytes for 
 $\mathcal{C}_1$ and pyramidal neurons for $\mathcal{C}_2$ . The values of the CDFs at the cosine similarities $\psi(\mathcal{G}, t)$,
 for clique labeled $\mathcal{G}$ and cell type labeled $t$, are plotted as colored circles. The values of the probability $\mathcal{P}_R(\mathcal{G}, t)$ are found in the third column of Table 1a,b (see Eq. \ref{estimateP}). The plots show that granule cells  and Purkinje cells both sit extremely comfortably at the top of the distribution of cosine similarities to the expresion of the clique.}
    \label{figureCDFTypes}
  \end{figure}

\begin{figure}[!ht]
\centering
    \subfloat[\label{subfig-1:dummy}]{%
      \includegraphics[width=0.9\textwidth]{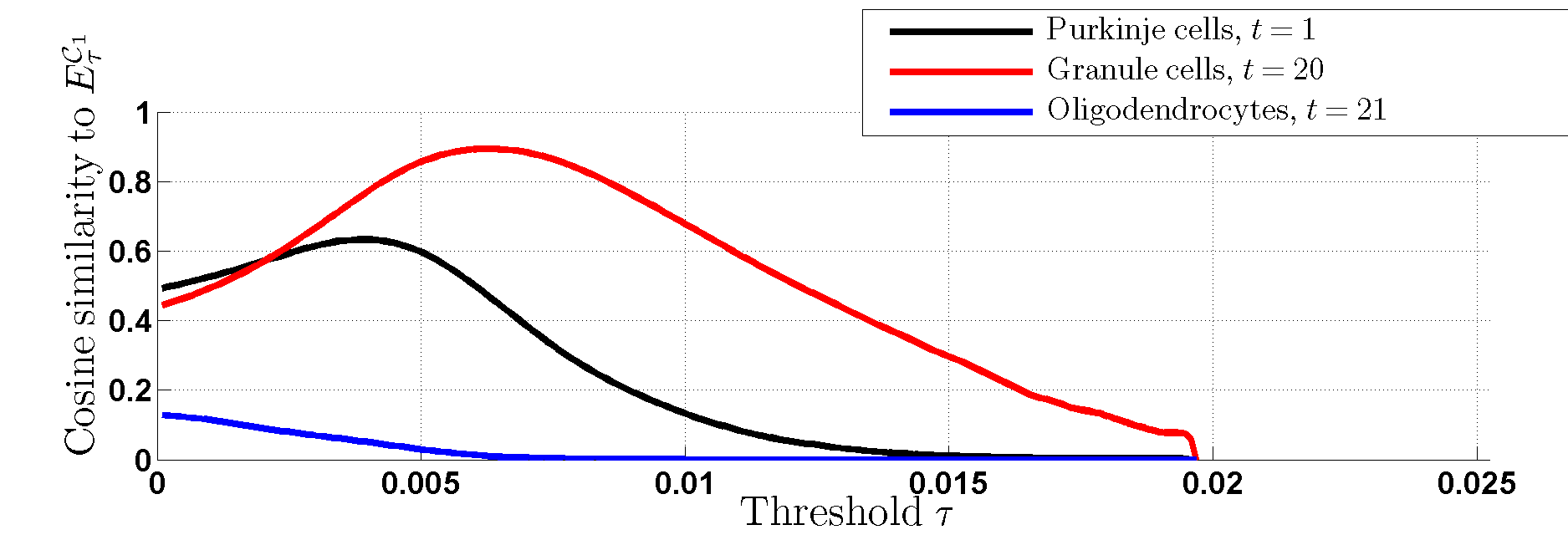}
    }
    \hfill
    \subfloat[\label{subfig-2:dummy}]{%
      \includegraphics[width=0.9\textwidth]{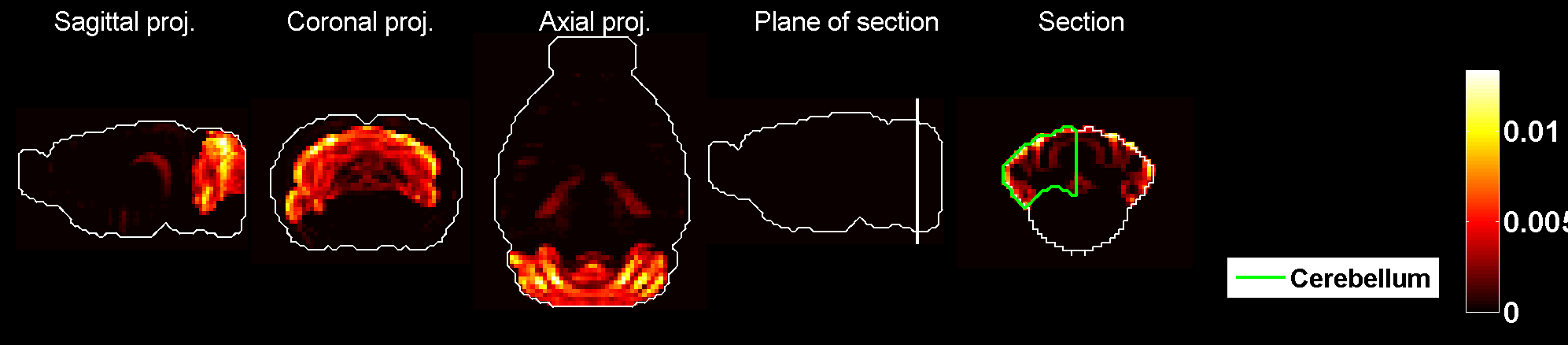}
    }
     \hfill
    \subfloat[\label{subfig-2:dummy}]{%
      \includegraphics[width=0.9\textwidth]{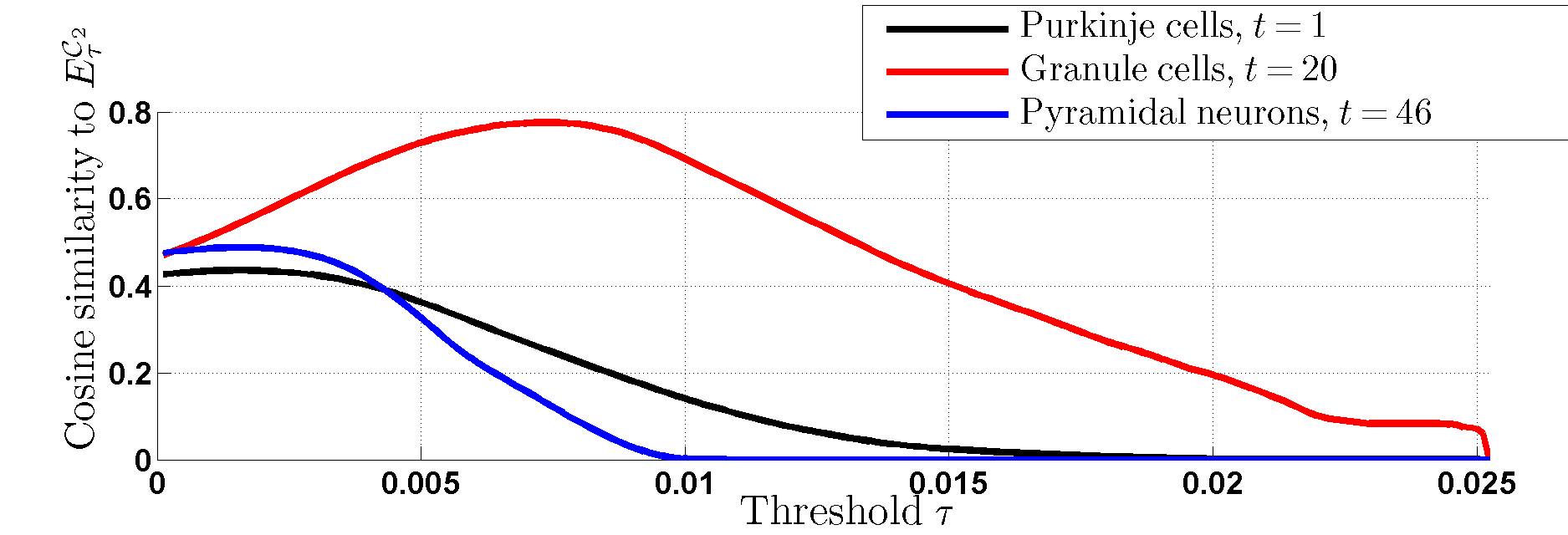}
    } 
  \hfill
    \subfloat[\label{subfig-2:dummy}]{%
      \includegraphics[width=0.9\textwidth]{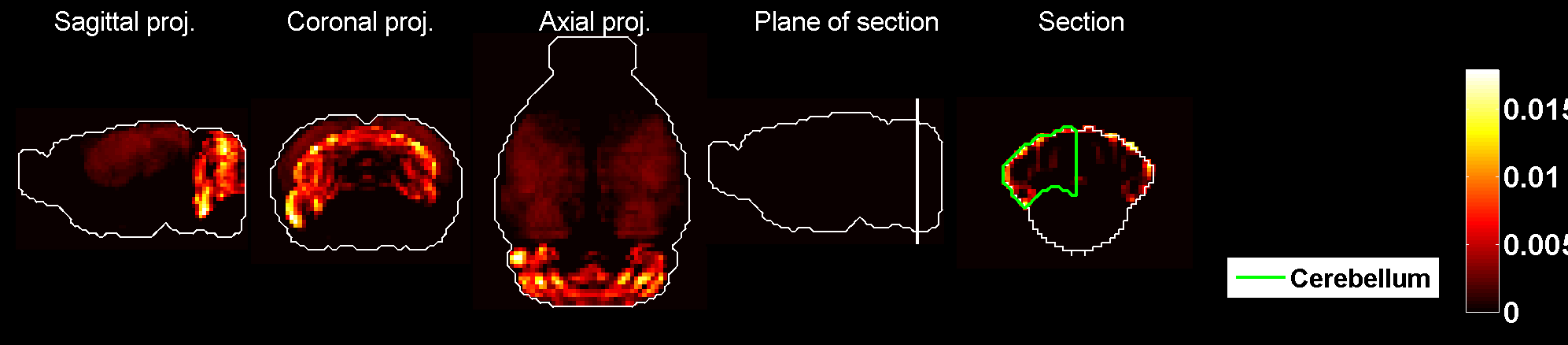}
    }
    \caption{Cosine similarities of thresolded gene expression energies of cliques, as 
 a function of the threshold (the expression profiles of the cliques are $L^2$-normalized so that threshold parameter $\tau$
 interpolates between the minimum and maximal value of each of them, and stays in the same range). (a) Plot of $\psi_\tau(\mathcal{C}_1,t)$
 as a function of $\tau$ for the top three cell types  in Table 1a. (c) Plot of $\psi_\tau(\mathcal{C}_1,t)$  as a function of $\tau$  
 for the top three cell types  in Table 1b. (b) Heat map of the expression energy of clique $\mathcal{C}_1$ at the value of the threshold $\tau$
 for which  $\psi_\tau(\mathcal{C}_1,20)$ is maximum. (d) Heat map of the expression energy of clique $\mathcal{C}_2$ at the value of the threshold $\tau$
 for which  $\psi_\tau(\mathcal{C}_2,20)$ is maximum.}
    \label{figContrast}
  \end{figure}

The cell-type-specific sample of granule cells, (labeled $t=20$) is the only  
 cell type that has a score higher than $99\%$ in both cliques. Figure 2 shows plots
  of the simulated 
 CDFs of the cosine similarities between the top three cell types by significance
 and sets of genes of the same size as ${\mathcal{C}}_1$ (Fig. \ref{figureCDFTypes}a) and ${\mathcal{C}}_2$ (Fig. \ref{figureCDFTypes}b).
   One can observe that both granule cells and
 Purkinje cells sit more comfortably at the top of the distribution than the cell type ranked third by statistical significance, especially 
 for clique $\mathcal{C}_2$.\\

 We therefore need to vary the contrast in the presentation of the expression patterns, in order to 
  decide in which sense, if any, the density profiles of granule cells and Purkinje cells are highlighted differently by 
 the cliques $\mathcal{C}_1$ and $\mathcal{C}_2$. We computed the cosine similarities between cell types
 and thresholded expression profiles of the two cliques, as defined by Eq. \ref{coExprThresh}.
  The values are plotted as a function of the threshold in Fig. \ref{figContrast}a,c.
 Granule cells present a peak for both cliques (Purkinje cells do only for the clique $\mathcal{C}_1$, but at a lower value 
 of the threshold and the top of the peak is lower,
 even though Purkinje cells start from a larger similarity to the clique $\mathcal{C}_1$  than granule cells before any threshold is applied).
  On the other hand, the thresholding procedure lowers the similarity between both cliques and the third
 cell type returned by the statistical analysis (oligodendrocytes for $\mathcal{C}_1$ and pyramidal neurons for $\mathcal{C}_2$.
 Moreover, Figure \ref{figContrast}b,d) shows heat maps of the expression profiles of both cliques,
 at the values corresponding to the peak of cosine similarity to granule cells. Indeed the coronal sections
 through the cerebellum exhibit the characteristic layered, hollow profile of the density of granule
cells observed in Fig. \ref{figHeatMaps}c2, which confirms that the granular layer is highlighted with more intensity 
 by the cliques than the Purkinje layer. Maximal-intensity projections of the thresholded expression profiles exhibit residual 
 expression in the cortex for clique $\mathcal{C}_2$, and to a lesser extent in the hippocampus
 for clique $\mathcal{C}_1$ (but it should be noted that genes are more highly expressed in the hippocampus than in any other region 
 of the brain on average in the coronal ABA).\\

We therefore conclude that the gene expression profiles of the two cliques of genes in this study
 highlight the cerebellum with more
 intensity in the granular layer than in the Purkinje layer, but these 
 two neuroanatomical structures are by far the most exceptionally similar to the expression
 profiles of the cliques.

\section{Discussion}

 Our computational analysis shows that among the cell types analyzed in \cite{cellTypeBased}, 
 the similarity of the expression of both cliques  to  granule cells and Purkinje cells is larger than 
 the similarity of more than 97\% of the cliques of the same size, and these two cell types 
 are the only cell types in the panel to have this property. The statistical significance of the similarity to the spatial density 
of granule cells is larger than the one of Purkinje cells for the clique ${\mathcal{C}}_2$, but 
  still Purkinje cells stand out together with granule cells (which makes sense with the involvement of Purkinje cells
 in autism discovered in post-mortem studies \cite{postMortem}). 
This completes our previous conclusion, reached in \cite{autismCoExpr} by visual inspection of 
 the Purkinje and granular layers of the cerebellar cortex. Granule cells (and not Purkinje cells) may 
 be present in some superficial voxels in which both cliques are highly expressed (see the coronal sections in \ref{figHeatMaps}), but as brain-wide neuroanatomical patterns,
  granule cells and Purkinje cells are both exceptionally similar to the expression profiles of 
  the two cliques in this study.\\

 The values of the cosine similarities are not ranked in the same order as 
 the statistical significances, 
because their values are not decreasing in the fourth columns of Tables 
 1a,b. This is related to the fact that
 the cosine similarity is biased in favor of cell types with a large support (and for example pyramidal neurons, $t=46$, have a larger support, at 8980 voxels,
  than granule cells, at 3351 voxels).  So, if a clique of genes has a large support (which is the case of both cliques in this study, which have 
non-zero expression at more than $49,000$ voxels), it can have a larger cosine similarity to pyramidal neurons than to granule cells, but its
 similarity to granule cells may be more statistically significant. This is the case for clique $\mathcal{C}_2$, 
and the fact is illustrated in more detail on Fig. 
 \ref{figureCDFTypes}b, where it is clear that the similarity between pyramidal neurons (labeled $t=46$) and clique $\mathcal{C}_2$, albeit larger 
 than the value for granule cells and Purkinje cells, sits lower in the distribution of the similarities. Our probabilistic 
 approach is therefore a useful complement to the computation of cosine similarities.\\

Our analysis shows that the gene-based approach of the ABA
 and the cell-based approach of the transcriptional classification of cell types in the brain 
 can be combined in order to quantify the similarity between expression patterns of 
 condition-related genes. Our results are limited by the paucity of the cell-type-specific data,
 since the number of transcriptionally disctinct neuronal cell types is presumably 
 much higher than $64$. However, the classification of cell types is a hierarchical problem, 
 and it is plausible that granule cells and Purkinje cells branch early from each other (and from cortical pyramidal neurons
and oligodendrocytes) 
 in the classification, which makes the available data set reasonably effective as a first draft in the context of this study.
  The computational methods we devised can be easily reapplied when more cell-type-specific microarray data become available.
 Moreover, alternative measures of similarity  can easily be substituted to the cosine similarity, without modifying 
 the analysis of statistical significance and contrast, or the number of random draws dictated by 
 Hoeffding's inequality. \\

The spatial resolution of the voxelized ISH data of the mouse ABA (200 microns) complicates the separation
 between granule cells and Purkinje cells, which we attempted here by our thresholding procedure,
  due to the extreme difference in size between the two cell types. Granule cells and Purkinje cells 
 may be present in the same voxel, and registration errors are therefore  much larger in scale 
 of a granule cell than in scale of a Purkinje cell. 
  An interesting analysis for such analysis can be found in \cite{KoCellTypes,JiCellTypes}, where image series rather than 
  voxelized data are used.\\

 The translation of results from the mouse model to humans is technically challenging,
 even though the ABA of the human brain has been released \cite{humanAtlas}, because the 
 human atlas cannot be voxelized, due to the size and paucity of the specimens. A first 
 step in this direction of research involves the study of variability of gene expression 
 between the mouse and human atlas in well-charted regions of the brain.

\end{document}